\documentclass[a4paper]{article}

\usepackage[numbers, sort]{natbib}

\usepackage{INTERSPEECH2022}
\usepackage{xspace}
\usepackage{algorithm,algorithmicx,algpseudocode}
\usepackage{mathtools}
\usepackage{hyperref}
\usepackage{url}
\usepackage{subcaption}
\usepackage{kotex}
\usepackage{multirow}
\usepackage{lscape}
\usepackage{graphicx}
\usepackage{tikz}
\usetikzlibrary{calc, shapes}
\usetikzlibrary{decorations.text}
\usetikzlibrary{arrows.meta}

\title{SANE-TTS: Stable And Natural End-to-End Multilingual Text-to-Speech}
\name{Hyunjae Cho$^{1, 2}$, Wonbin Jung$^{1, 3}$, Junhyeok Lee$^{1}$, Sang Hoon Woo$^{1}$}
\address{
  $^1$MINDsLab Inc., Republic of Korea\\
  $^2$Seoul National University, Republic of Korea\\
  $^3$Korea Advanced Institute of Science
and Technology (KAIST), Republic of Korea}
\email{\{chohyunjae, wbjung, jun3518, shwoo\}@mindslab.ai}

\makeatletter
\DeclareRobustCommand\onedot{\futurelet\@let@token\@onedot}
\def\@onedot{\ifx\@let@token.\else.\null\fi\xspace}

\makeatother

\newcommand{\mltts}{multilingual TTS\xspace}
\newcommand{\themodel}{SANE-TTS\xspace}

\begin{document}

\maketitle

\begin{abstract}
  In this paper, we present SANE-TTS, a stable and natural end-to-end multilingual TTS model.
By the difficulty of obtaining multilingual corpus for given speaker, training multilingual TTS model with monolingual corpora is unavoidable.
We introduce speaker regularization loss that improves speech naturalness during cross-lingual synthesis as well as domain adversarial training, which is applied in other multilingual TTS models.
Furthermore, by adding speaker regularization loss, replacing speaker embedding with zero vector in duration predictor stabilizes cross-lingual inference.
With this replacement, our model generates speeches with moderate rhythm regardless of source speaker in cross-lingual synthesis.
In MOS evaluation, SANE-TTS achieves naturalness score above 3.80 both in cross-lingual and intralingual synthesis, where the ground truth score is 3.99.
Also, SANE-TTS maintains speaker similarity close to that of ground truth even in cross-lingual inference.
Audio samples are available on our web page\footnote{\url{https://mindslab-ai.github.io/sane-tts/}}.
\end{abstract}

\section{Introduction} \label{section:introduction}
  While almost all works in text-to-speech (TTS) focus on synthesizing speech in a single language, \mltts aims to generate speeches in multiple languages with a single model.
  The most naive approach is training a model with multilingual speech dataset for a given speaker, but it is infeasible due to the unavailability of such datasets.
  Thus, prior works \cite{code_switching, meta_learning_tts, google_adversarial, cross_lingual_voice_cloning, cross_lingual_spk_emb} have focused on using a mix of monolingual corpora to build \mltts models by implementing cross-lingual speech synthesis.
  
  Previous \mltts models \cite{code_switching, meta_learning_tts, google_adversarial, cross_lingual_voice_cloning, cross_lingual_spk_emb} are mainly based on Tacotron \cite{tacotron, tacotron2}.
  However, Tacotron-based models have several issues while they synthesize speeches in an autoregressive manner that utilizes attention \cite{attention_mechanism} to align input text and target speech.
  First, attention errors cause wrong alignment estimation, resulting in the problem of words skipping and repeating \cite{deep_voice_3}.
  Second, generating results autoregressively with attention inhibits direct control of phoneme-level duration \cite{fastspeech_1}.
  On the other hand, some \mltts models \cite{yourtts, multilingual_tts_cross_lingual_vc} are not Tacotron-based models.
  YourTTS \cite{yourtts} focuses more on zero-shot learning and does not support cross-lingual synthesis.
  \citet{multilingual_tts_cross_lingual_vc} also proposed a \mltts model based on voice conversion.
  However, this study is closer to TTS data augmentation by the pre-trained voice conversion model and only covers the Indo-European languages with the International Phonetic Alphabet representation.
  
  Most non-autoregressive models \cite{glowtts, fastspeech_1, fastspeech_2, fastpitch, fastpitchformant, VITS} include duration predictor to estimate each phoneme's duration and total length without autoregressive iteration.
  Among the duration predictor-based models \cite{glowtts, fastspeech_1, fastspeech_2, fastpitch, fastpitchformant, meta_stylespeech, VITS, parallel_tacotron_2}, we choose VITS \cite{VITS}, an end-to-end TTS model, as a backbone.
  Furthermore, VITS is advantageous because it generates natural speeches with high synthesis speed and requires a single model training since it is not a two-stage model.
  While utilizing a duration predictor, the key challenge for multilingual TTS is the uncertainty of duration prediction in cross-lingual synthesis.
  
  In this paper, we propose \emph{\themodel}, a \mltts model with natural speech synthesis and stable duration prediction.
  Due to using monolingual corpora, speaker identity and linguistic features can be entangled.
  To generate natural speeches in cross-lingual synthesis, speaker identity needs to disentangle from linguistic features.
  So, we add \emph{speaker regularization loss} term to prevent language information leakage to speaker representation.
  To predict duration stably, duration predictor should produce phoneme duration independent of speaker identity during cross-lingual inference.
  For this reason, our duration predictor uses zero vector instead of speaker embedding generating moderate phoneme duration regardless of speaker identity.
  Our contributions can be summarized as:
  \begin{itemize}
    \item \emph{\themodel} synthesizes multilingual speeches stably with a perceptual score close to the ground truth level while maintaining speaker similarity even in cross-lingual inference.
    \item Proposed \emph{speaker regularization loss} achieves improvement of speech naturalness as much as domain adversarial training (DAT) \cite{dann}, which is commonly used in previous cross-lingual synthesis studies \cite{google_adversarial, meta_learning_tts}.
  \end{itemize}
\section{Method}
\begin{figure*}[!hbt]
  \centering
  \begin{subfigure}[b]{0.5\textwidth}
    \includegraphics[width=.9\columnwidth]{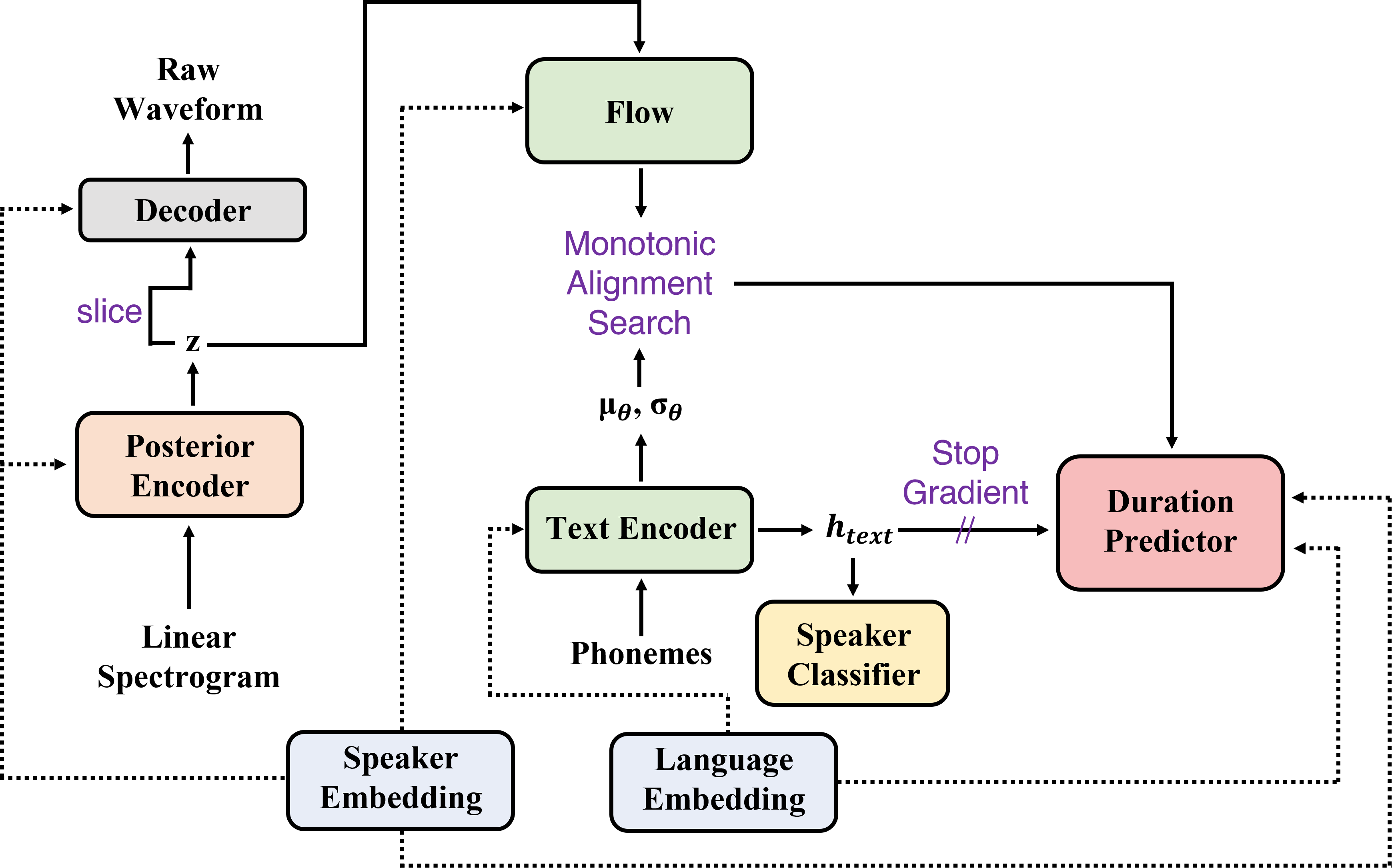}
    \caption{Training procedure}
    \label{fig:training}
  \end{subfigure}
  \begin{subfigure}[b]{0.49\textwidth}
    \includegraphics[width=.9\columnwidth]{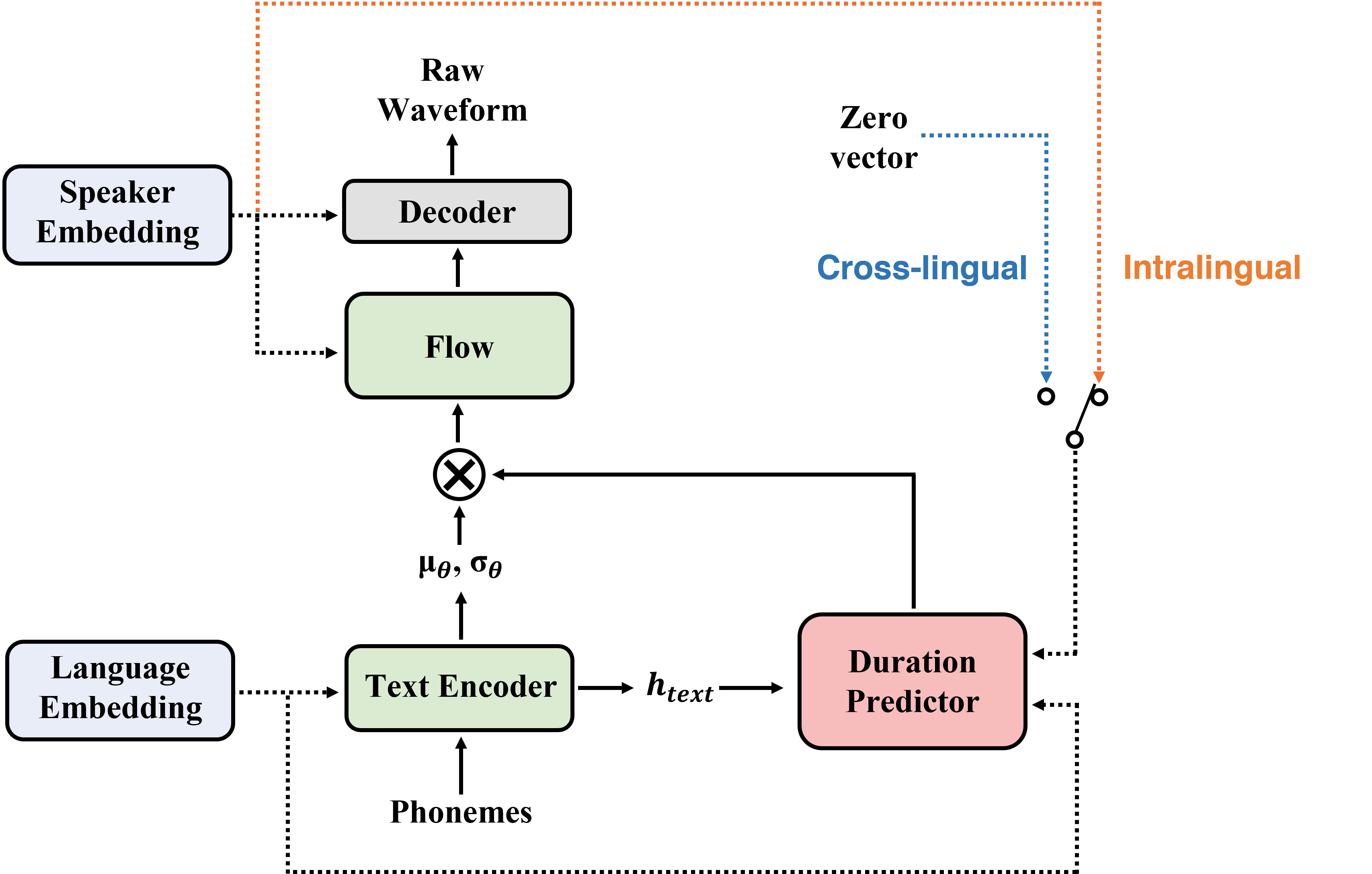}
    \caption{Inference procedure}
    \label{fig:inference}
  \end{subfigure}
  \caption{Block diagram of system overview of proposed model in (a) training procedure and (b) inference procedure.}
  \label{fig:overview}
\end{figure*}
  We modify some modules and loss terms to build \mltts model.
  To receive various languages, we change text encoder and duration predictor.
  For loss function, we apply DAT \cite{dann} to make text representation disentangle from speaker identity.
  Also, we add a speaker regularization loss term to learn language-independent speaker representations.
  Figure \ref{fig:overview} illustrates overview of our system during training procedure and inference procedure.
  \themodel gets phoneme sequences, speaker embedding, and language embedding, as inputs, and generates raw waveform as an output.
  We use different phoneme sets and grapheme-to-phoneme converters for each language during conversion of transcripts into phoneme sequences.
  In training procedure, posterior encoder gets linear spectrogram as an additional input.
  
\subsection{Language embedding}
  \themodel includes learnable language embeddings, which are 256-dimensional vectors, the same size with the speaker embeddings.
  We only provide language embedding to text encoder and duration predictor, which directly get text representation as an input.
  To provide language information in duration predictor, language embedding is passed through a convolution layer and added to the hidden text representation in the same manner as speaker embedding.
  For the flow, the posterior encoder, and the decoder, we retain the same setup as the original VITS.
  
\subsection{Text encoder}
  We use a Transformer-based text encoder \cite{transformer} with relative positional representation \cite{relative_positional_representation} same as VITS, with parameter generation suggested by \citet{meta_learning_tts} to be compatible with various languages.
  The parameter generator takes language embedding as an input and generates parameters of the text encoder.
  This process helps optimizing the text encoder for each language of input phoneme sequences.
  
\subsection{Domain adversarial training}
  Since every speaker does not share the same transcript, text representation in TTS model can be entangled with speaker identity.
  To reduce speakers' bias on the text representation, we use DAT \cite{dann, google_adversarial}.
  We attach a speaker classifier at the end of the text encoder as a domain classifier.
  The speaker classifier consists of fully connected layers, and a gradient reversal layer is inserted between the speaker classifier and the text encoder.
  We train the speaker classifier with cross-entropy loss to prevent predicting speaker identity from text representation.
  Through DAT, the text encoder learns speaker-independent text representation, and the model can generate speeches from general texts.
  
\subsection{Speaker regularization loss}
  Similar to speaker's bias on the text representation, speaker identity is biased by the language of the speaker's utterances in the dataset.
  To synthesize speech across various languages, we have to prevent speaker identities from entangling to languages in the duration predictor.
  To reduce speaker bias for language, we introduce a \emph{speaker regularization loss} $L_\mathrm{reg}$ given by:
  \begin{align}
    L_\mathrm{reg} &= \lVert\mathbb{E}_{k \in K}\left[\mathrm{conv}(S_{k})\right]\lVert_2,
    \label{reg_loss}
  \end{align}
  where $\mathrm{conv}$ is a convolution layer with a kernel size of 1 and $S_{k}$ represents the speaker embedding of a speaker in datapoint $k$ in the batch $K$.
  As the mean of hidden speaker representations $\mathrm{conv}(S_k)$ is pushed to zero vector regardless of the language, the speaker identities disentangle to languages in the model.
  
  Since the proposed speaker regularization loss pushes the mean of hidden speaker representations toward zero vector, the duration predictor estimates moderate phoneme duration by inputting a zero vector instead of speaker embedding in cross-lingual inference.
  On the other hand, the duration predictor gets speaker embedding as an input during intralingual inference because input text consists of seen phonemes by the speaker.
  This method reduces the instability of the duration predictor and removes the uncertainty of adjusting the speaker information to phoneme duration in cross-lingual synthesis.
  
\subsection{Deterministic duration predictor}
  Originally in VITS, a stochastic duration predictor (SDP) was proposed to generate speeches with diverse rhythms.
  SDP predicts phoneme duration stochastically from noise latent by normalizing flow \cite{flow}.
  However, \citet{yourtts} reported that there are cases where the SDP generates unnatural duration causing unclear pronunciation.
  So, in this paper, a deterministic duration predictor (DDP) \cite{glowtts, VITS} is applied to improve the stability of speech synthesis.
\section{Experiments}
\subsection{Dataset}
  We construct our dataset by gathering internal and external speech corpora \cite{ljspeech, libritts, kss, jvs, aishell} in four languages.
  The dataset is composed of speeches from multiple speakers in English (EN), Korean (KO), Japanese (JA) and Mandarin Chinese (ZH) as shown in Table \ref{tab:dataset_detail}.
  We resample audio samples to 22.05 kHz.
  We convert transcripts of speeches into phoneme sequences through our internal grapheme-to-phoneme conversion process.
  We hold out 5\% of utterances for validation set.

\begin{table}[th]
  \centering
  \caption{Details of our dataset in multiple languages}
  \label{tab:dataset_detail}
  \resizebox{1.0\linewidth}{!}{
    \begin{tabular}{l|r r r r|r}  
      \toprule
      Language              & EN        & KO        & JA        & ZH        & Total     \\
      \midrule
      Number of speakers    & 161       & 27        & 110       & 174       & 472       \\
      Length                & 72.4hr    & 43.5hr    & 27.5hr    & 60.0hr    & 203.4 hr  \\
      \bottomrule
    \end{tabular}
  }
\end{table}
  
\subsection{Setup}
  We train our model for 200 epochs on 2 NVIDIA A100 GPUs.
  We use mixed precision training with batch size of 64.
  We follow the schedule of scale factor $\lambda$ which weights the speaker classification loss in DAT \cite{dann} as:
  \begin{align}
    \lambda &= \frac{2}{1 + \exp(-10 \cdot p)} - 1,
    \label{schedule_lambda}
  \end{align}
  where $p$ is the training progress linearly changing with the training steps, from 0 to 1.
  The loss scale factor is initiated at 0 and continues to increase as training progresses.
  This schedule suppresses DAT at the early stage and improves the quality of outputs.
  Other details follow that of VITS \cite{VITS}.

\subsection{Baseline model} \label{sec:exp_comparison}
  To demonstrate that our model generates high-quality speeches during cross-lingual inference, we compare \themodel with an official open-sourced implementation of another multilingual TTS model (meta-learning model)\footnote{\url{https://github.com/Tomiinek/Multilingual\_Text\_to\_Speech}} proposed by \citet{meta_learning_tts}.
  We train meta-learning model up to 100 epochs with a batch size of 64.
  For vocoding, we use the Griffin-Lim algorithm \cite{gla} provided in the official implementation. 
  
 \subsection{Ablation study} \label{sec:exp_ablation}
  We compare \themodel with models that remove a single modification from our model.
  We remove three modifications; applying DAT, adding speaker regularization loss, and replacing SDP by DDP.
  When removing our speaker regularization loss, we input speaker embedding in the duration predictor during cross-lingual inference.
  We exclude modifications regarding language embedding, which is necessary to implement our \mltts model.
  
\subsection{Evaluation}
  We report the mean opinion scores (MOS) to evaluate the naturalness of speeches and the similarity of the speaker between speech pairs of ground truth and synthesized samples \cite{sim_mos}, including 95\% confidence intervals.
  Our MOS evaluation uses an absolute category rating scale, where raters score performances from 1 to 5 in 1 point increments.
  We conduct MOS evaluation on Amazon Mechanical Turk framework.
  Due to the difficulty of acquiring native speakers for rating non-English target languages, we gather opinion scores on English speeches from native English speakers.
  Each speech and speech pair is scored by 5 raters.
  For speaker similarity, raters compare a synthesized speech with a speech of the same speaker in the validation set and evaluate speaker similarity between them.
  In the case of ground truth, we select two speeches of the same speaker in the validation set and raters evaluate their speaker similarity.
  
  To generate evaluation speech samples, we sample 30 sentences randomly from the English validation set.
  Also, we select 5 male and 5 female speakers from every four languages to evaluate both the intralingual and the cross-lingual synthesis capabilities.
  We synthesize speeches for every combination of speakers and sentences, total 1200 utterances.
\section{Results and Discussion}
\subsection{Speech synthesis quality} \label{sec:result_comparison}
  Table \ref{tab:comparison_naturalness} and Table \ref{tab:comparison_similarity} show naturalness MOS and similarity MOS of the comparative models with 95\% confidence intervals.
  \themodel surpasses meta-learning model both in naturalness and speaker similarity.
  Our model achieves naturalness MOS of 3.95 for the intralingual synthesis which is close to that of the ground truth 3.99.
  Also, naturalness of the cross-lingual synthesis does not change significantly in the intralingual synthesis, above 3.80 MOS.
  For both naturalness MOS and similarity MOS, English speech samples get higher scores than other languages.
  It is because of different levels of difficulty of intralingual and cross-lingual synthesis.
  Also, the evaluation could be biased by the raters who are English speakers.

\begin{table}[th]
  \centering
  \caption{Comparison of naturalness MOS with another model}
  \label{tab:comparison_naturalness}
  \resizebox{1.0\linewidth}{!}{
  \begin{tabular}{lcccc}
    \toprule
    \multicolumn{1}{c}{\multirow{2}{*}{Model}} & \multicolumn{1}{c|}{Intralingual} & \multicolumn{3}{c}{Cross-lingual} \\ \cmidrule{2-5}
    \multicolumn{1}{c}{} & \multicolumn{1}{c|}{EN} & KO & JA & ZH \\
    \midrule
    Ground truth & 3.99 $\pm$ 0.04 & - & - & - \\
    \midrule
    \themodel& \textbf{3.95 $\pm$ 0.04} & \textbf{3.80 $\pm$ 0.05} & \textbf{3.84 $\pm$ 0.04} & \textbf{3.81 $\pm$ 0.04} \\
    Meta-learning model & 3.19 $\pm$ 0.06 & 3.30 $\pm$ 0.06 & 3.03 $\pm$ 0.06 & 2.97 $\pm$ 0.06 \\
    \bottomrule
  \end{tabular}
  }
\end{table}

\begin{table}[th]
  \centering
  \caption{Comparison of similarity MOS with another model}
  \label{tab:comparison_similarity}
  \resizebox{1.0\linewidth}{!}{
  \begin{tabular}{lcccc}
    \toprule
    \multicolumn{1}{c}{\multirow{2}{*}{Model}} & \multicolumn{1}{c|}{Intralingual} & \multicolumn{3}{c}{Cross-lingual} \\ \cmidrule{2-5}
    \multicolumn{1}{c}{} & \multicolumn{1}{c|}{EN} & KO & JA & ZH \\
    \midrule
    Ground truth & 3.38 $\pm$ 0.05 & 3.60 $\pm$ 0.05 & 3.44 $\pm$ 0.05 & 3.50 $\pm$ 0.05 \\
    \midrule
    \themodel & \textbf{3.48 $\pm$ 0.06} & \textbf{3.31 $\pm$ 0.06} & \textbf{3.26 $\pm$ 0.06} & \textbf{3.44 $\pm$ 0.06} \\
    Meta-learning model & 2.92 $\pm$ 0.06 & 2.81 $\pm$ 0.07 & 2.59 $\pm$ 0.06 & 2.73 $\pm$ 0.06 \\
    \bottomrule
  \end{tabular}
  }
\end{table}

\subsubsection{The cross-lingual speech synthesis}
\newcommand{\incaptionimg}[3]{
  \begin{tikzpicture}[every node/.style={inner sep=0,outer sep=0}]
    \draw node[name=micrograph] {\includegraphics[width=\columnwidth]{#1}}; 
    \draw  (micrograph.north west)  node[anchor=north west,yshift=-0.15cm,xshift=0.35cm,#3]{\small{(#2)}}; 
  \end{tikzpicture}
}
\begin{figure}[t]
  \centering
  \vspace{-1pt}
  \incaptionimg{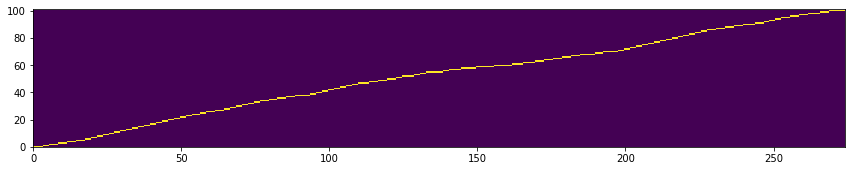}{a}{white}
  \vspace{-1pt}
  \incaptionimg{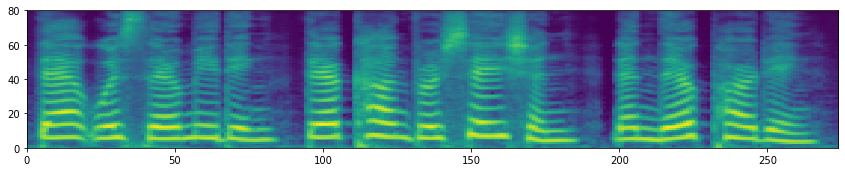}{b}{white}
  \vspace{-1pt}
  \incaptionimg{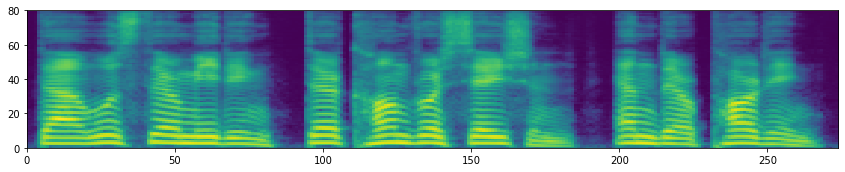}{c}{white}
  \vspace{-1pt}
  \incaptionimg{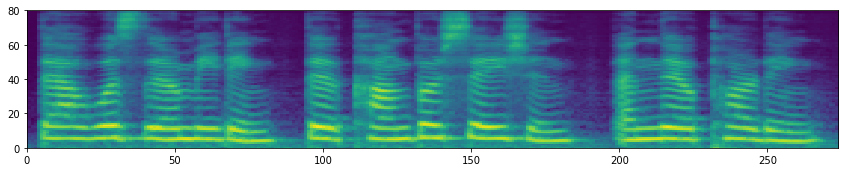}{d}{white}
  \vspace{-2pt}
  \caption{Cross-lingual alignment and mel spectrograms of the text ``That was their meeting, their conversation, and their parting." (a) is cross-lingual alignment between the text and speech. (b-d) are from synthesized speech of (b) KSS (KO) \cite{kss}, (c) jvs001 (JA) \cite{jvs}, and (d) SSB0018 (ZH) \cite{aishell}.}
  \label{fig:stable_alignment}
\end{figure}
  Figure \ref{fig:stable_alignment} shows an alignment and mel spectrograms of speeches synthesized by \themodel during cross-lingual inference.
  Our model generates identical alignment for every speaker since DDP receives a zero vector instead of speaker embedding.
  Thus, \themodel synthesizes speeches with moderate rhythm regardless of source speaker.
  
\subsection{Ablation study} \label{sec:result_ablation}
  Table \ref{tab:ablation_naturalness} and Table \ref{tab:ablation_similarity} show naturalness MOS and similarity MOS of the ablation study with 95\% confidence intervals.
  \themodel shows degradation of naturalness MOS in the cross-lingual inference without the proposed speaker regularization loss upto 0.11 and DAT upto 0.06.
  Specifically, removing speaker regularization loss decreases speech naturalness slightly more than excluding DAT in cross-lingual synthesis.
  
\begin{figure}[t]
  \centering
  \vspace{-1pt}
  \incaptionimg{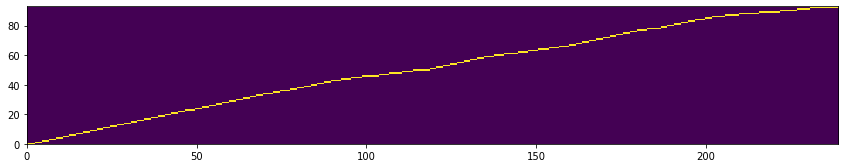}{a}{white}
  \vspace{-1pt}
  \incaptionimg{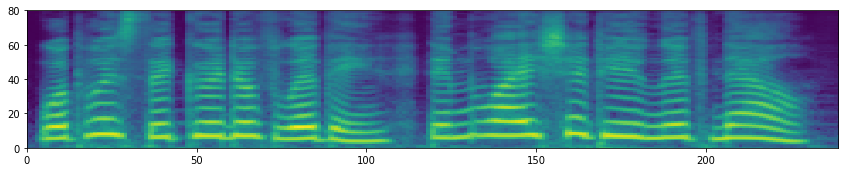}{b}{white}
  \vspace{-1pt}
  \incaptionimg{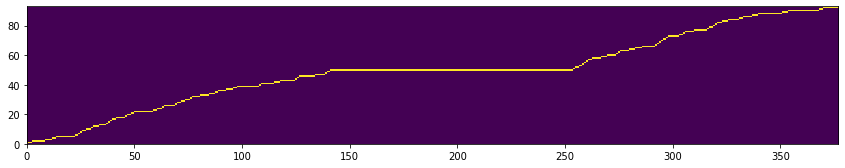}{c}{white}
  \vspace{-1pt}
  \incaptionimg{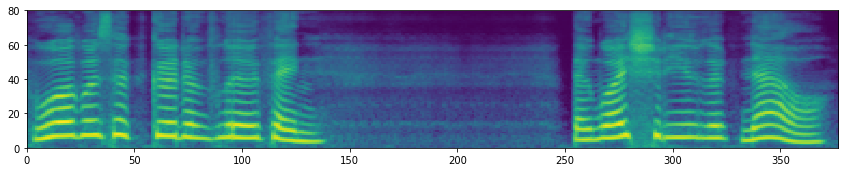}{d}{white}
  \vspace{-1pt}
  \caption{Alignments and mel spectrograms of the speech generated by models with DDP and SDP. The text of the speech is ``What was the good of living, and why should he live now?" and the speaker is KSS \cite{kss}. The model with DDP generates (a) alignment and (b) mel spectrogram. In extreme cases, model with SDP generates (c) alignment and (d) mel spectrogram with overlong silence.}
    \label{fig:sdp and ddp alignment and mel-spectrogram}
\end{figure}

  There is no statistically significant difference between the ablation model with SDP and our model.
  However, there are cases where SDP predicts an unnatural duration in cross-lingual synthesis.
  Figure \ref{fig:sdp and ddp alignment and mel-spectrogram} depicts generated alignments and mel spectrograms from models with DDP and SDP.
  The SDP predicts unnatural duration that produces an overlong silence in the middle of the speech.
  Although SDP produces speeches with diverse rhythms, we apply relatively stable DDP for reliable \mltts model.
  
\begin{table}[th]
  \centering
  \caption{Comparison of naturalness MOS in the ablation study}
  \label{tab:ablation_naturalness}
  \resizebox{1.0\linewidth}{!}{
  \begin{tabular}{lcccc}
    \toprule
    \multicolumn{1}{c}{\multirow{2}{*}{Model}} & \multicolumn{1}{c|}{Intralingual} & \multicolumn{3}{c}{Cross-lingual} \\ \cmidrule{2-5}
    \multicolumn{1}{c}{} & \multicolumn{1}{c|}{EN} & KO & JA & ZH \\
    \midrule
    Ground truth        & 3.99 $\pm$ 0.04 & - & - & - \\
    \midrule
    \themodel            & \textbf{3.95 $\pm$ 0.04} & \textbf{3.80 $\pm$ 0.05} & \textbf{3.84 $\pm$ 0.04} & \textbf{3.81 $\pm$ 0.04} \\
    w/o DAT             & 3.85 $\pm$ 0.05 & 3.77 $\pm$ 0.04 & 3.82 $\pm$ 0.04 & 3.75 $\pm$ 0.04 \\
    w/o Regularization  & 3.88 $\pm$ 0.04 & 3.69 $\pm$ 0.05 & 3.76 $\pm$ 0.04 & 3.72 $\pm$ 0.05 \\
    w/ SDP              & 3.93 $\pm$ 0.04 & \textbf{3.81 $\pm$ 0.05} & 3.80 $\pm$ 0.04 & 3.74 $\pm$ 0.05 \\
    \bottomrule
  \end{tabular}
  }
\end{table}

\begin{table}[th]
  \centering
  \caption{Comparison of similarity MOS in the ablation study}
  \label{tab:ablation_similarity}
  \resizebox{1.0\linewidth}{!}{
  \begin{tabular}{lcccc}
    \toprule
    \multicolumn{1}{c}{\multirow{2}{*}{Model}} & \multicolumn{1}{c|}{Intralingual} & \multicolumn{3}{c}{Cross-lingual} \\ \cmidrule{2-5}
    \multicolumn{1}{c}{} & \multicolumn{1}{c|}{EN} & KO & JA & ZH \\
    \midrule
    Ground truth        & 3.38 $\pm$ 0.05 & 3.60 $\pm$ 0.05 & 3.44 $\pm$ 0.05 & 3.50 $\pm$ 0.05 \\
    \midrule
    \themodel            & \textbf{3.48 $\pm$ 0.06} & \textbf{3.31 $\pm$ 0.06} & \textbf{3.26 $\pm$ 0.06} & \textbf{3.44 $\pm$ 0.06} \\
    w/o DAT             & 3.33 $\pm$ 0.06 & 3.16 $\pm$ 0.06 & 3.15 $\pm$ 0.06 & 3.40 $\pm$ 0.06 \\
    w/o Regularization  & 3.34 $\pm$ 0.06 & 3.29 $\pm$ 0.06 & 3.18 $\pm$ 0.06 & 3.47 $\pm$ 0.06 \\
    w/ SDP              & 3.45 $\pm$ 0.06 & \textbf{3.34 $\pm$ 0.06} & 3.26 $\pm$ 0.06 & \textbf{3.48 $\pm$ 0.06} \\
    \bottomrule
  \end{tabular}
  }
\end{table}

\subsubsection{Visualizing regularization of speaker identities}
\begin{figure}[th]
  \centering
  \includegraphics[width=\columnwidth]{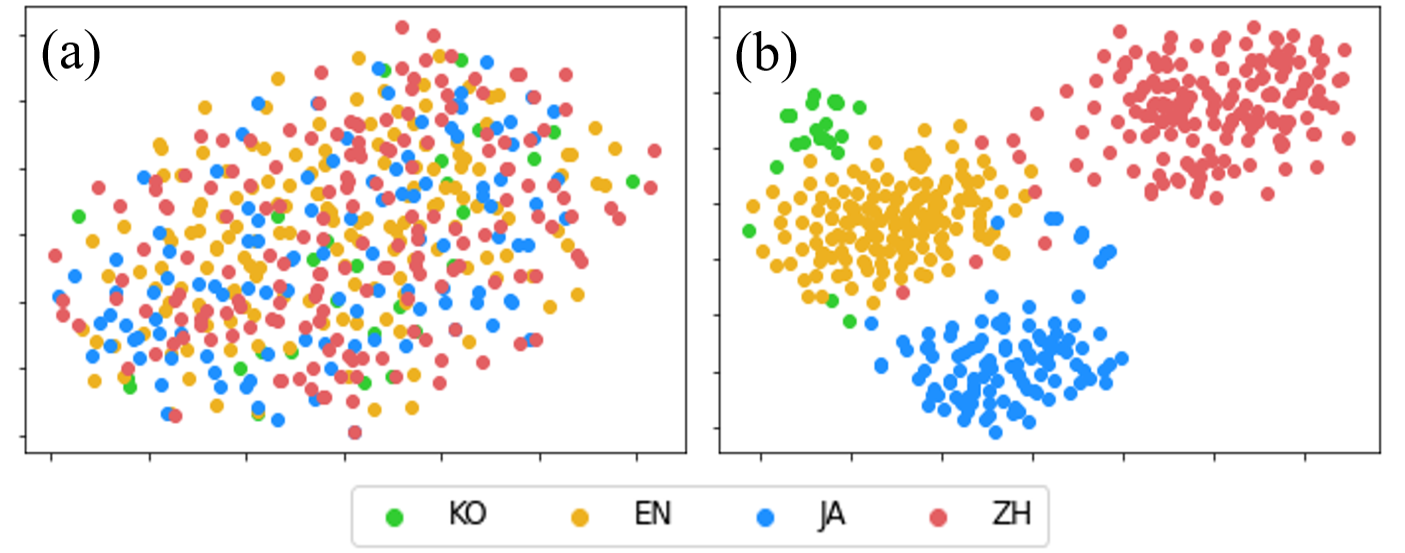}
  \vspace{-1pt}
  \caption{t-SNE plots of hidden speaker representations for the duration predictor for every 472 speakers (a) with speaker regularization loss and (b) without speaker regularization loss.}
  \label{fig:t-SNE}
\end{figure}
  Figure \ref{fig:t-SNE} shows the t-SNE plots of hidden speaker representations for the duration predictor with and without the speaker regularization loss.
  Without the speaker regularization loss, the hidden speaker representations form clusters by languages, while no clusters are observed, and samples are distributed around the center with the speaker regularization loss.
  The t-SNE plots demonstrate that our model learns language-independent speaker representations by adding the speaker regularization loss.
\section{Conclusions}
  In this paper, we propose \themodel, a stable and natural end-to-end \mltts model.
  Due to the limited multilingual speech corpus, we use a mix of monolingual corpora for training \mltts model.
  Therefore, \mltts model faces the difficulty of cross-lingual inference for languages that are not recorded by target speaker.
  To solve difficulty of \mltts, we introduce \emph{speaker regularization loss} to make our model learns speaker representation independently from its own language.
  Also, by replacing speaker embedding with zero vector in the cross-lingual duration prediction, the model produces moderate phoneme duration irrelevant to speaker identity.
  In addition, we add language embedding and apply DAT which are commonly used techniques.
  In our multilingual setup with English, Korean, Japanese and Mandarin Chinese, \themodel generates natural audio samples which obtain high speaker similarity during both the cross-lingual and the intralingual synthesis.
  Based on our study, we expect to expand \themodel into other languages in future work.
\section{Acknowledgements}
  The authors would like to thank Seungu Han and Kang-wook Kim from MINDsLab Inc., Jinwoo Kim from KAIST, and Hyeongkeun Kim from University of Illinois Urbana-Champaign for providing beneficial feedback on this work.

\bibliographystyle{IEEEtranN}
\bibliography{main}

\end{document}